# A scale of interpretation for likelihood ratios and Bayes factors

Frank Dudbridge

*Department of Population Health Sciences, University of Leicester, Leicester, UK*

Department of Population Health Sciences, University of Leicester, University Road, Leicester LE1 7RH, UK

frank.dudbridge@leicester.ac.uk

Orcid ID 0000-0002-8817-8908

According to the odds version of Bayes' theorem

$$\frac{\Pr(H_0|X)}{\Pr(H_1|X)} = \frac{\Pr(X|H_0)}{\Pr(X|H_1)} \frac{\Pr(H_0)}{\Pr(H_1)}$$

the likelihood ratio is additive on the log-odds of $H_0$ to $H_1$ (I will use "likelihood ratio" to refer to both simple likelihood ratios and Bayes factors). The likelihood ratio is on an absolute continuous scale, but in practice its precise value may be less important than its order of magnitude. This is recognised in fields that weigh the evidence for simple hypotheses. For example forensic science has adopted a scaling with base 10 (1). In clinical diagnostics, a rule of thumb is used in which likelihood ratios of 2, 5 and 10 correspond to risk differences of 15%, 30% and 45% (2).

In Bayesian testing too, several authors have suggested descriptors of evidential strength. Jeffreys (3) proposed a scaling with base 10, specifically 1 to $10^{\frac{1}{2}} \approx 3.2$ ("bare mention"), 3.2 to 10 ("substantial"), 10 to $10^{\frac{3}{2}} \approx 32$ ("strong"), 32 to $10^2 = 100$ ("very strong") and $> 100$ ("decisive"). (The hypotheses are ordered such that the likelihood ratio >1). A variation was given by Kass and Raftery (4) using natural logs: $\exp(1) \approx 3$ ("bare mention"), $\exp(3) \approx 20$ ("positive"), $\exp(5) \approx 150$ ("strong"). For Goodman (5), up to 5 is "weak", 10 "moderate", 20 "strong" and 100 "very strong".

Royall (6) proposed a scaling with base 2 motivated by a "canonical urn experiment". We draw balls from an urn with replacement, comparing the hypothesis that all the balls are black with the alternative that half are black, half are white. Each black ball drawn doubles the likelihood ratio in favour of all black balls. Of course there is no reason why there should only be two colours of ball, nor why the alternative is an equal proportion. Furthermore, a single white ball voids all previously accrued evidence. All of that aside, Royall considered three black balls to be "fairly strong" evidence (likelihood ratio 8) and 5 black balls to be "strong" (likelihood ratio 32).

The various proposals are compared in Table 1. Obviously there is variation in how certain values are described, and in how much evidence qualifies as "strong". This reflects the subjectivity of the authors, each of whom cited their personal experience. But for practitioners it is unclear which scaling is most appropriate for general use.

| Likelihood ratio | Jeffreys (3) | Kass and Raftery (4) | Royall (6) | Goodman (5) |
|---|---|---|---|---|
| 3 | Bare mention | Bare mention | | |
| 5 | | | | Weak |
| 8 | | | Fairly strong | |
| 10 | Substantial | | | Moderate |
| 20 | | Positive | | Strong |
| 32 | Strong | | Strong | |
| 100 | Very strong | | | Very strong |
| >100 | Decisive | | | |
| 150 | | Strong | | |
| >150 | | Very strong | | |

**Table 1. Scaling of evidence, rounded to integers, proposed by different authors. Entries show the descriptor applied to a likelihood ratio less than or equal to the corresponding row label and greater than that of the previous entry in the same column. Hypotheses are ordered such that the likelihood ratio >1.**

Note also that the use of English language may lend itself to cultural bias, and evolving connotation. The scales in Table 1 were proposed in the 20[th] century, predating a surge in high volume data; what was "strong" then may not seem so today, let alone in 500 years. Indeed Lee and Wagenmakers (7) and Held and Ott (8) have already renewed Jeffreys' thresholds with different adjectives.

There seems then to be a demand for a qualitative scale of evidence, but no objective scale is yet available. In the following proposal, a unit of evidence is defined to be the likelihood ratio that updates weaker belief to stronger belief, where those terms are given a quantitative definition in terms of the effect of evidence on belief. This definition leads automatically to a sequence of qualitative levels that may be extended indefinitely.

Suppose that inferences are drawn on the probability (not the odds) of $H_0$ given the choice between $H_0$ and $H_1$. Then the probability $p$ is related to the log odds $l$ by the logistic function

$$p = u(l) = \left(1 + e^{-l}\right)^{-1}$$

Since the likelihood ratio is additive on the log odds scale, the first derivative represents the *effect of evidence on probability*. As shown in Figure 1, and easily verified by calculus, evidence has the strongest effect when $p = ½$, and the least effect at $p = 0$ and $p = 1$. When probability has a belief interpretation, it is easy to think of examples from everyday life that reflect this relationship.

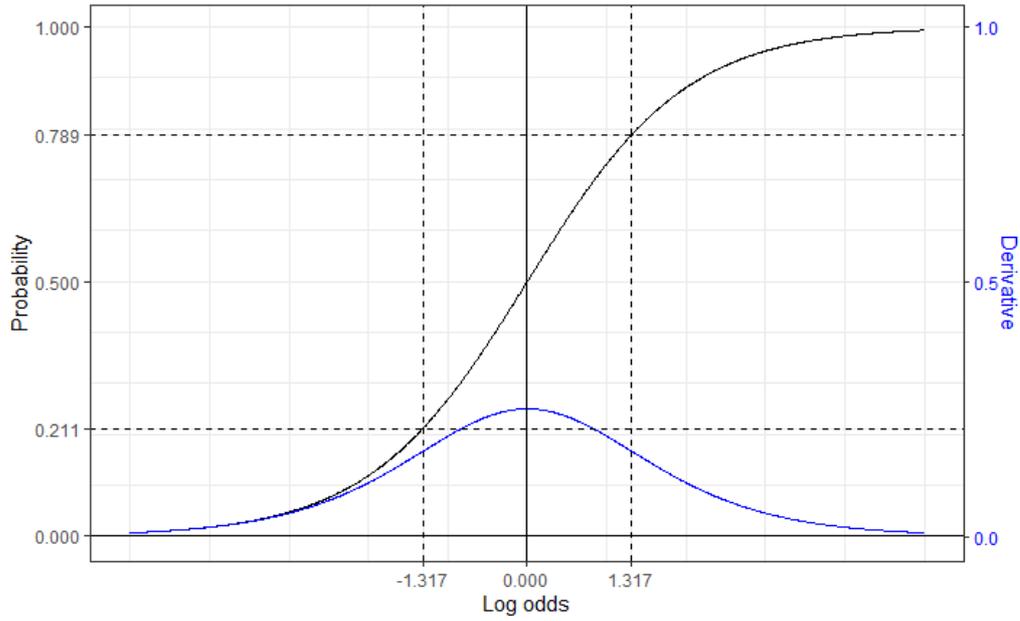

**Figure 1: Evidence and probability.** Relationship between log odds and probability (logistic function, upper black curve, left axis) and the effect of evidence on probability (first derivative, lower blue curve, right axis). Dashed lines indicate the sharpest change in the effect of evidence, defining one unit.

We can distinguish between probabilities for which the effect of evidence is higher, from those where the effect of evidence is lower. There is no hard threshold since the effect of evidence varies continuously, but the transition from higher to lower effects is sharpest at the extrema of the second derivative, that is the zeros of the third. The derivatives of the logistic function can be expressed as polynomials in the function itself (9):

$$u^{(n)}(\cdot) = (-1)^n \sum_{k=0}^{n-1} \left\langle {n \atop k} \right\rangle u^{(k+1)}(u-1)^{n-k}$$

where $\left\langle {n \atop k} \right\rangle$ are called Eulerian numbers (10) and are computed as

$$\left\langle {n \atop k} \right\rangle = \sum_{j=0}^{k} (-1)^j \binom{n+1}{j}(k-j+1)^n$$

The third derivative is zero at $l = \pm \log\left(\frac{\sqrt{3}+1}{\sqrt{3}-1}\right) = \pm 1.317$ with $p = \frac{1}{2}\left(1 \pm \frac{1}{\sqrt{3}}\right) = \{0.211, 0.789\}$ (9). This is where the sharpest distinction can be made between higher and lower effects of evidence. Let us correspondingly describe probability in the interval (0.5, 0.789] as weaker belief, and (0.789, 1] as stronger belief. Similarly describe [0.211, 0.5) as weaker disbelief, and [0, 0.211) as stronger disbelief. Now define one unit of evidence as the minimum required to update any weaker belief to stronger belief.

**Definition 1.** *Likelihood ratio $L$ comprises $\log_b L$ units of evidence, where* $b = \inf\left\{x : x\theta > \frac{\sqrt{3}+1}{\sqrt{3}-1}\right\}$ *for all $\theta > 1$. Therefore* $b = \frac{\sqrt{3}+1}{\sqrt{3}-1} = 3.73$.

Presented with likelihood ratio $L$, a simple interpretation is that it entails evidence equivalent to $(\log_b L)$ independent studies that each just update weaker belief to stronger belief. How one interprets "weaker" and "stronger" remains subjective, but note the comparative forms of those terms.

An extended interpretation can be constructed as follows. From definition 1, one unit of evidence would also update weaker disbelief to weaker belief, and two units would update weaker disbelief to stronger belief. Having so defined one and two units, then if the prior probability is weaker *belief*, two units would update it to the interval (0.933, 1]. Call this "much stronger belief", say. Then the minimum evidence to update any weaker *disbelief* to much stronger belief is three units. Now having defined three units, we can define another interval ("much, much stronger belief") by updating weaker belief by that amount. This process can be extended indefinitely, giving an interpretation of any amount of evidence in relation to any prior probability (Table 2). This interpretation avoids any absolute description of the evidence itself.

|  | Weaker belief | Stronger belief | Much stronger belief | Much, much stronger belief |
|---|---|---|---|---|
| Weaker belief |  | 1 | 2 | 3 |
| Weaker disbelief | 1 | 2 | 3 | 4 |
| Stronger disbelief | 2 | 3 | 4 | 5 |
| Much stronger disbelief | 3 | 4 | 5 | 6 |
| Much, much stronger disbelief | 4 | 5 | 6 | 7 |

**Table 2. Units of evidence that update prior probabilities (rows) to posterior probabilities (columns). "Weaker" and "stronger" are defined by the arguments supporting definition 1, giving the entries in the top left 2x2 sub-table. Further categories can be added as described in the text. For simplicity they are understood as disjoint in the table, so for example "stronger belief" is the probability range (0.789, 0.933].**

Table 3 shows the likelihood ratios for the first four units of evidence. They are similar to the subjective levels proposed previously (Table 1), and suggest approximate rules of thumb. Also shown are the posterior probabilities when the prior probability is ½. This gives support to colloquial statements such as "95% certain", but in principle the units of evidence stand independently of prior probability.

| Units | Likelihood ratio | Rule of thumb | Probability | Rule of thumb |
|---|---|---|---|---|
| **1.0** | 3.73 | 4 | 0.789 | 80% |
| **2.0** | 13.9 | 15 | 0.933 | 95% |
| **3.0** | 52.0 | 50 | 0.981 | 98% |
| **4.0** | 194 | 200 | 0.995 | 99% |

**Table 3. Likelihood ratios corresponding to the first four units of evidence. Probability, posterior probability when the prior probability is ½.**

Units of evidence allow us to discard English-language descriptors for more objective levels. The argument stems from the idea that a weak belief is one that is easily changed by evidence, and a strong belief is one that is not easily changed. Ideally there would be a threshold at which the effect of evidence changes abruptly from high to low. However, the logistic relationship is imposed by Bayes' theorem and we must make do with a softer threshold defined by the sharpest change in the continuous effect of evidence. Of course this threshold should not be taken as a hard rule, but as a guide to the qualitative interpretation of the likelihood ratio. Instead of describing the evidence as "moderate" or "substantial", we can talk about one or two units of evidence, and develop an intuition about the contextual import of such evidence.

The units can also aid in the specification of prior probabilities. It can be hard, and often unnecessary, to be precise about either the prior probability or the required level of posterior probability. Arguably, if a hypothesis is being tested at all, there must be *prima facie* evidence in its favour. Even prior odds may represent one limit, with the first few units providing other values for consideration. In general, the scale with base 3.73 suggests that likelihood ratios within a factor of 4 are at a similar level of evidence.